\documentclass[useAMS,usenatbib]{mn2e}

\usepackage{times}
\usepackage{graphicx}
\usepackage{subfig}

\title[Asteroseismology of KUV 11370+4222]{Asteroseismology of the ZZ Ceti star KUV 11370+4222}
\author[J. Su et al.]{J. Su,$^{1,2,3}$\thanks{E-mail:sujie@ynao.ac.cn} Y. Li,$^{1,2}$ J.-N. Fu,$^{4}$ and C. Li$^{4}$ \\
$^1$Yunnan Astronomical Observatory, Chinese Academy of Sciences, P.O. Box 110, Kunming 650011, China \\
$^2$Key Laboratory for the Structure and Evolution of Celestial Objects, Chinese Academy of Sciences \\
$^3$University of Chinese Academy of Sciences, Beijing 100049, China \\
$^4$Department of Astronomy, Beijing Normal University, Beijing 100875, China}

\begin{document}

\date{Accepted 2013 October 23. Received 2013 October 23; in original form 2013 September 11}

\pagerange{\pageref{firstpage}--\pageref{lastpage}}
\pubyear{}

\maketitle

\label{firstpage}

\begin{abstract}
KUV 11370+4222 is a ZZ Ceti star discovered in 1996, which has not been observed since then. We performed observations for KUV 11370+4222 in 2010 January. From the Fourier transform spectrum of the light curves, ten independent modes are detected. We searched for the best-fitting model by using the observed periods to match the model periods and get it with the total mass of 0.625\,${\rm M_{\sun}}$, the effective temperature of 10950\,K, and the helium mass fraction and hydrogen mass fraction of $10^{-2.2}$ and $10^{-4.0}$, respectively. We have found a triplet of frequency split by rotation, which are $l=1$ modes. Using the frequency shifts we estimate a rotation period of $5.56\pm0.08$\,h. Besides it, another $l=1$ mode is identified. The other observed periods are identified as $l=2$ modes. At last we investigated the property of the mode trapping.
\end{abstract}


\section{Introduction}

White dwarf stars are in the final stage of the stellar evolution. About 98 per cent of all stars \citep{wd08} will become white dwarf stars in the end. They provide us important samples to investigate the late evolutionary stage of stars. White dwarf stars are also ready-made laboratories for testing physics under extreme conditions. The majority (about 80 per cent) of all white dwarfs are of DA type, who have hydrogen-rich atmospheres. There is an observational instability strip for DA white dwarfs in the effective temperature ($T_{\rm eff}$) range of 12270 to 10850\,K \citep[see][and references therein]{cb07}. When the DA white dwarfs evolve and pass through the instability strip they become pulsators. These pulsating DA white dwarfs are known as the DAV or ZZ Ceti stars. Arlo Landolt discovered the first DA pulsator HL Tau 76 \citep{la68}. Up to now, the total number of ZZ Ceti stars has risen to 148 \citep{cb10}.

The oscillations observed in the pulsating white dwarfs are driven by the $\kappa$-mechanism due to the partial ionization of hydrogen at the blue edge of the instability strip and by convective driving at cooler temperature. The buoyancy is the restoring force and these oscillating modes are called g-modes. Since different oscillation modes propagate through different zones inside the star, they provide information about the invisible interiors of stars. By detecting and analysing these modes, one can probe the internal structure of the pulsating stars. The technique is called asteroseismology, somewhat similar to the technique that seismologists use to study the interior of the Earth by probing the earthquake waves.

The fact that the DA instability strip is pure \citep*{ga11} indicates that going across the instability strip is a necessary evolutionary phase of DA white dwarfs, which all have the opportunities to become ZZ Ceti stars. Hence the internal structure of the ZZ Ceti stars investigated by asteroseismology is representative of the entire DA white dwarfs.

Asteroseismology relies on theoretical models of white dwarfs as well. The first step of building a model is to derive a number of essential parameters of the star, such as the total mass, the effective temperature, the hydrogen mass fraction and the helium mass fraction. Asteroseismology is a unique way to determine the mass fractions of hydrogen and helium, and to give more accurate values of the stars' masses and the effective temperatures, compared with the technique of spectrometry.

KUV 11370+4222 was originally discovered to be a ZZ Ceti star by \citet{vg97}. No observations had been carried out since then. Three modes with the periods of 257.2, 292.2 and 462.9\,s were detected in the short observation run, which are insufficient to determine neither the stellar parameters nor the internal structure of KUV 11370+4222. We thus decided to make observations for KUV 11370+4222 using the newly-built 2.4-m telescope of Yunnan Astronomical Observatory in China. Our goal was to collect more data to make a preliminary asteroseismological study for KUV 11370+4222.

In this paper, we introduce the observations and the model analysis for KUV 11370+4222. The observations and data reduction are described in Section \ref{observations}. In Section \ref{fourier}, we give the results of Fourier analysis with the extracted signals. In Section \ref{asteroseismology}, we present the details of the asteroseismological study for KUV 11370+4222. The introduction of theoretical tools is given in Section \ref{tools}; the description of model matching in Section \ref{matching}; the results of the best-fitting model in Section \ref{best-fitting}; the mode identification in Section \ref{identification}; the rotation period of the star in Section \ref{rotation} and the discussion about mode trapping in Section \ref{trapping}. At last we give the summary and conclusions in Section \ref{summary}.

\section{Observations and data reduction} \label{observations}

KUV 11370+4222 was observed for eight nights from 2010 January 24 to 31. The observations were performed with the 2.4-m telescope located at Lijiang station of Yunnan Astronomical Observatory in China. The images of stars were taken with a Princeton Instruments VersArray:1300B back-illuminated CCD camera. The field of view of the camera is about $4.5\times4.5$\,arcmin$^2$. Johnson $B$ filter was used during the observations. The exposure time of each frame was 40\,s. In order to make the readout noise as low as possible, we chose the low speed mode to read each frame out. Meanwhile, we set the CCD camera binned $2\times2$ to reduce the sampling time. The journal of observations is listed in Table~\ref{journal}. The duration of this observation run is about 173.4\,h with a duty cycle of 23 per cent, which leads to a frequency resolution of 1.6\,$\umu$Hz.

\begin{table*}
\centering
\caption{Journal of observations for KUV 11370+4222 in 2010 January. Time is given in Heliocentric Julian Date (HJD).}
\begin{tabular}{cccccc}
\hline
Date & Start time    & End time      & Hours & Frame number \\
     & (HJD-2455200) & (HJD-2455200) &       &              \\
\hline
24 & 21.252 & 21.470 & 5.23 & 410 \\
25 & 22.271 & 22.386 & 2.76 & 189 \\
26 & 23.262 & 23.468 & 4.94 & 375 \\
27 & 24.153 & 24.476 & 7.75 & 602 \\
28 & 25.248 & 25.395 & 3.53 & 267 \\
29 & 26.251 & 26.478 & 5.45 & 427 \\
30 & 27.262 & 27.471 & 5.02 & 394 \\
31 & 28.255 & 28.477 & 5.33 & 417 \\
\hline
\end{tabular}
\label{journal}
\end{table*}

\begin{figure*}
\centering
\includegraphics{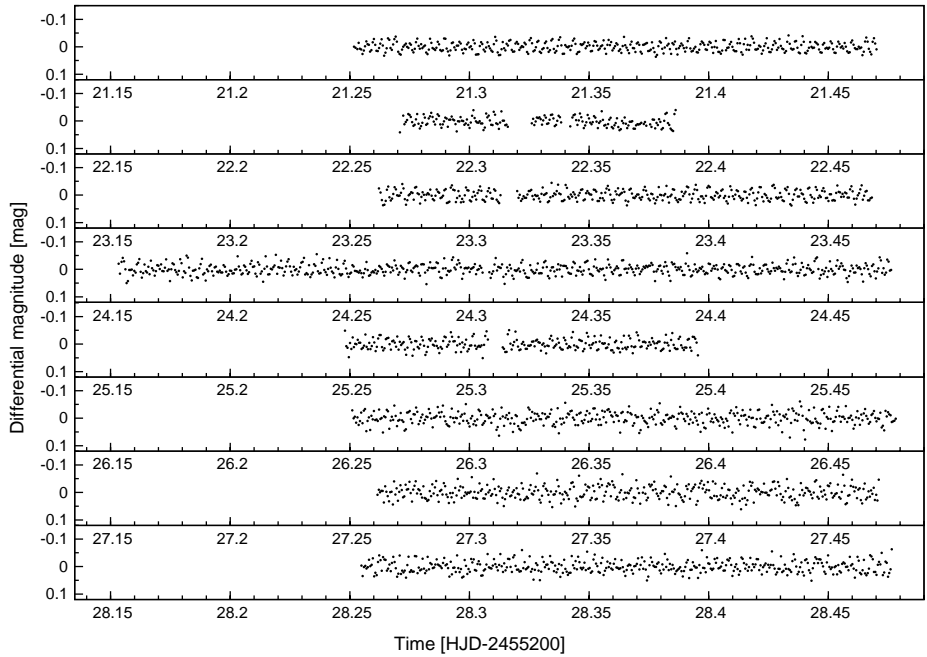}
\caption{Light curves of KUV 11370+4222 observed from 2010 January 24 to 31. The abscissa is the time in HJD$-$2455200. The ordinate is the differential magnitude of the object star relative to the comparison star.}
\label{lc}
\end{figure*}

The data were reduced with the {\small IRAF} routines. During the pre-reduction, bias and flat were corrected for all frames. The dark correction was ignored since the CCD camera was operating at $-$120\,$^{\circ}$C with liquid nitrogen cooling hence the dark current was less than 1\,e\,p$^{-1}$\,h$^{-1}$ in the condition. We employed the {\small IRAF APPHOT} package to perform aperture photometry. In order to optimize the sizes of the aperture, we used different apertures for the data in each night and took the aperture which brought the minimum variance of the light curves of the check star relative to the comparison star. The data reduction was carried out with the standard process of aperture photometry. The light curves of KUV 11370+4222 relative to the comparison star were obtained. In order to filter the low frequency signals caused by the variation of the atmospheric transparency during the night, the light curves were divided by a fifth order polynomial. At last, the time of each data point was converted to the Heliocentric Julian Date (HJD). The light curves are shown in Fig.~\ref{lc}.

\section{Fourier analysis} \label{fourier}

\begin{figure*}
\centering
\includegraphics{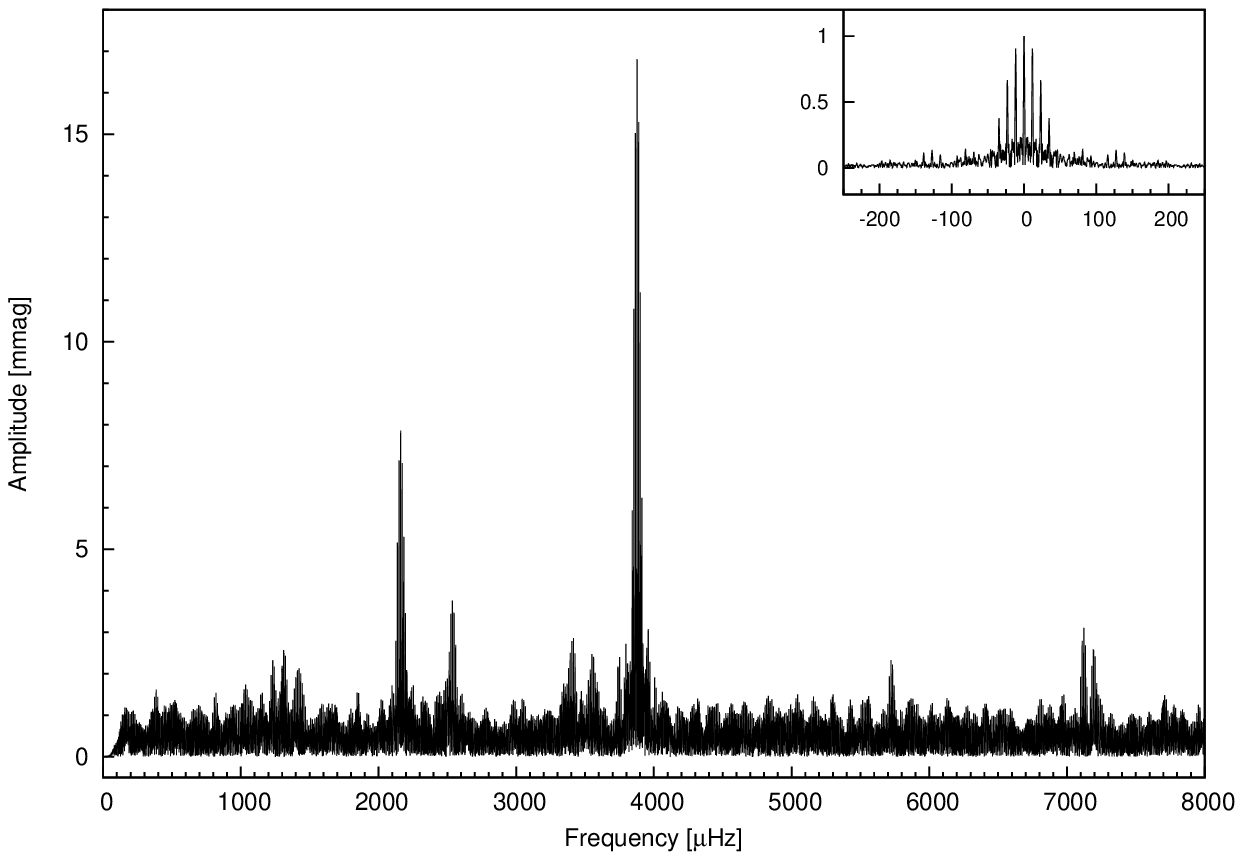}
\caption{Amplitude spectrum of the light curves. The frequency is in microhertz ($\umu$Hz) and the amplitude in milli-magnitude (mmag). We only show the spectrum with frequency range from 0 to 8000\,$\umu$Hz, since no significant signal is found outside that range. The inset shows the spectral window.}
\label{ft}
\end{figure*}

We use the software {\small PERIOD04} \citep{lp05} to perform Fourier transform (FT) of the light curves. {\small PERIOD04} was designed to analyse multi-periodic astronomical time series, allowing analysis of the data containing gaps. The amplitude spectrum is shown in Fig.~\ref{ft}.

\subsection{Extraction of pulsation signals}

We followed a standard prewhitening procedure to extract the frequencies, amplitudes and phases of the pulsation signals from the FT spectrum:
\begin{enumerate}
\item Pick out the frequency of the highest peak in the original FT spectrum.
\item Construct a sinusoid with the obtained frequency and derive the amplitude and phase by fitting the sinusoid to the observed light curves.
\item Subtract the sinusoid from the light curves.
\item Get the FT spectrum of the residual light curves.
\item Pick out the frequency of the highest peak in the residual FT spectrum as the next frequency.
\item Construct a new sinusoid with the frequency, and fit the residual light curves to get the corresponding amplitude and phase.
\item Repeat steps (iii) through (vi) to obtain the next signals.
\end{enumerate} 
An open question is where is the end of the iteration. We adopt a widespread criterion \citep{bm93,kr97} that signals whose level of signal-to-noise ratio (S/N) are less than 4 are probably not reliable. So if the newly extracted signal has a value of S/N\,$<$\,4, we refuse it and stop the procedure. We finally get twelve signals with S/N\,$>$\,4 from the FT spectrum, which are listed in Table~\ref{signals}. The uncertainties of the frequency and amplitude were estimated with a Monte Carlo Simulation, as done by \citet{fj13}.

\begin{table*}
\centering
\caption{The S/N\,$>$\,4 signals extracted from the FT spectrum, listed according to the frequency values from low to high.}
\begin{tabular}{cccrr}
\hline
ID & Frequency  & Period & Amplitude & S/N \\
   & ($\umu$Hz) & (s)    & (mmag)    &     \\
\hline
$f_{01}$ & 1231.91 $\pm$ 0.19 & 811.75 &  2.27 $\pm$ 0.50 &  4.02 \\
$f_{02}$ & 1311.62 $\pm$ 0.15 & 762.42 &  2.61 $\pm$ 0.53 &  4.41 \\
$f_{03}$ & 2160.62 $\pm$ 0.05 & 462.83 &  7.76 $\pm$ 0.48 & 13.09 \\
$f_{04}$ & 2487.19 $\pm$ 0.23 & 402.06 &  2.21 $\pm$ 0.53 &  4.38 \\
$f_{05}$ & 2511.97 $\pm$ 0.21 & 398.09 &  2.33 $\pm$ 0.52 &  4.51 \\
$f_{06}$ & 2537.14 $\pm$ 0.10 & 394.14 &  4.10 $\pm$ 0.56 &  6.88 \\
$f_{07}$ & 3403.99 $\pm$ 0.17 & 293.77 &  2.72 $\pm$ 0.51 &  5.08 \\
$f_{08}$ & 3561.21 $\pm$ 0.18 & 280.80 &  2.48 $\pm$ 0.52 &  4.94 \\
$f_{09}$ & 3877.73 $\pm$ 0.02 & 257.88 & 16.87 $\pm$ 0.46 & 18.64 \\
$f_{10}$ & 5721.88 $\pm$ 0.19 & 174.77 &  2.33 $\pm$ 0.51 &  4.45 \\
$f_{11}$ & 7122.21 $\pm$ 0.13 & 140.41 &  3.00 $\pm$ 0.50 &  6.34 \\
$f_{12}$ & 7192.56 $\pm$ 0.19 & 139.03 &  2.40 $\pm$ 0.53 &  5.05 \\
\hline
\end{tabular}
\label{signals}
\end{table*}

In Table~\ref{compare}, the amplitude of the three periods found in 1996 is listed. We compare them to the corresponding ones of 2010 in the same table. It shows remarkable amplitude variations, which are similar to many other ZZ Ceti stars. The mechanism producing those variations is yet unknown.

\begin{table*}
\centering
\caption{Comparison of the amplitude ($A$) of the three periods ($P$) found in 1996 with the corresponding ones found in 2010.}
\begin{tabular}{cccc}
\hline
\multicolumn{2}{c}{1996} & \multicolumn{2}{c}{2010} \\
$P$ (s) & $A$ (mmag) & $P$ (s) & $A$ (mmag) \\
\hline
462.90 & 3.20 & 462.83 &  7.76 \\
292.20 & 2.50 & 293.77 &  2.72 \\
257.20 & 5.30 & 257.88 & 16.87 \\
\hline
\end{tabular}
\label{compare}
\end{table*}

\subsection{Harmonics and linear combinations}

Some of the signals are probably harmonics or linear combinations of the pulsation modes. Only the real eigen modes can be used for asteroseismology. Suppose there are three signals A, B and C, whose frequencies (with uncertainties) are $f_{\rm A} \pm \sigma_{\rm A}$, $f_{\rm B} \pm \sigma_{\rm B}$ and $f_{\rm C} \pm \sigma_{\rm C}$, respectively. In case of linear combinations, the frequencies are related by the linear relation among them
\[
(f_{\rm A} \pm \sigma_{\rm A}) \approx (f_{\rm B} \pm \sigma_{\rm B}) \pm (f_{\rm C} \pm \sigma_{\rm C}).
\]
We take the signals with the higher amplitudes as the independent modes and the lower-amplitude one as the combination. Further more, the possible linear combination should satisfy the criterion that
\[
\sigma_{\rm A} \le \sigma_{\rm B}+\sigma_{\rm C}.
\]
 
Checking the signals, we find that
\begin{enumerate}
\item $f_{10} \approx f_{03} + f_{08}$, which means $f_{10}$ (5721.88 $\pm$ 0.19\,$\umu$Hz) is likely the linear combination of $f_{03}$ (2160.62 $\pm$ 0.05\,$\umu$Hz) and $f_{08}$ (3561.21 $\pm$ 0.18\,$\umu$Hz).  Note that the amplitude of $f_{10}$ is smaller than the amplitude of both $f_{03}$ and $f_{08}$.
\item $f_{11} \approx 2f_{08}$, which indicates $f_{11}$ (7122.21 $\pm$ 0.13\,$\umu$Hz) is probably a harmonic of $f_{08}$ (3561.21 $\pm$ 0.18\,$\umu$Hz).
\end{enumerate}
The uncertainties of $f_{10}$ and $f_{11}$ are $ \sigma_{10} = 0.19$\,$\umu$Hz $ < \sigma_{03} + \sigma_{08} = 0.23$\,$\umu$Hz and $\sigma_{11} = 0.13$\,$\umu$Hz $ < 2\sigma_{08} = 0.36$\,$\umu$Hz, respectively, which fulfil the criterion.

Generally speaking, larger amplitude modes are likely to form combinations. It is surprising that $f_{03}$ is the second largest amplitude peak which combines with $f_{08}$ to form a linear combination $f_{10}$, while the largest amplitude peak $f_{09}$ does not form any combination.

After excluding $f_{10}$ and $f_{11}$, ten frequencies remained to be considered as independent modes. We list the values of frequency and period of these independent modes in Table~\ref{modes}. The differences between the adjacent frequencies and periods are also calculated and listed in the table.

\begin{table*}
\centering
\caption{A list of independent modes. The 1st column is the frequencies ($f$) and the 2nd column frequency differences ($\Delta f$) between adjacent frequencies. The 3rd and the 4th columns list the corresponding periods ($P$) and period differences ($\Delta P$).}
\begin{tabular}{cccc}
\hline
$f$ ($\umu$Hz) &                       & $P$ (s) &                \\
               & $\Delta f$ ($\umu$Hz) &         & $\Delta P$ (s) \\
\hline
1231.91 &         & 811.75 &        \\
        &   79.71 &        &  49.33 \\    
1311.62 &         & 762.42 &        \\
        &  849.00 &        & 299.59 \\ 
2160.62 &         & 462.83 &        \\
        &  326.56 &        &  60.77 \\
2487.19 &         & 402.06 &        \\
        &   24.78 &        &   3.97 \\ 
2511.97 &         & 398.09 &        \\
        &   25.17 &        &   3.95 \\  
2537.14 &         & 394.14 &        \\
        &  866.85 &        & 100.37 \\ 
3403.99 &         & 293.77 &        \\
        &  157.22 &        &  12.97 \\
3561.21 &         & 280.80 &        \\
        &  316.52 &        &  22.92 \\
3877.73 &         & 257.88 &        \\
        & 3314.83 &        & 118.85 \\  
7192.56 &         & 139.03 &        \\
\hline
\end{tabular}
\label{modes}
\end{table*}

\subsection{A triplet} \label{triplet}

It is known that the g-modes in white dwarfs are non-radial oscillations. An individual oscillation mode is characterized by three integer numbers: $l$, $n$ and $m$. The spherical harmonic degree $l$ indicates the total number of nodal lines on the spherical surface. The radial order $n$ denotes the number of nodes of the radial eigenfunction. The azimuthal order $m$ indicates how many of these nodal lines cross the equator, and $m$ can only be chosen as the integer values which satisfy $|m| \le l$. If a star is spherically symmetric, the azimuthal order $m$ would be degenerate, i.e. modes with different values of $m$ will have exactly the same frequency. The existence of rotation will break the star's spherical symmetry and remove the degeneracy of $m$. This effect would split a frequency into $2l+1$ components ($|m| \le l$), which brings triplets for $l=1$ modes, quintuplets for $l=2$ modes, septuplets for $l=3$ modes, and so on.

One can find that the three modes: 2487.19, 2511.97 and 2537.14\,$\umu$Hz form a triplet, whose frequency spacing is of $\sim$ 25\,$\umu$Hz (see Table~\ref{modes}). It thus offers information for us to identify the three modes as rotational splitting of $l=1$ mode. The central frequency 2511.97 \,$\umu$Hz is of $m=0$ mode, and the two companions (2487.19 and 2537.14\,$\umu$Hz) are of $m=-1$ and $m=1$ respectively.

\section{Asteroseismology of KUV 11370+4222} \label{asteroseismology}

An important step of asteroseismology is to match the pulsation periods of theoretical models to the observed periods in order to find the best-fitting model.

\subsection{Theoretical tools} \label{tools}

To get the theoretical models, we employ the White Dwarf Evolution Code (WDEC) with some modifications. WDEC was originally written by Martin Schwarzschild \citep[see][for more details]{sm65} and modified subsequently by other researchers \citep{kg69,ld75,wd81,ks86,wm90,bp93,mm98,mt01}. The evolution begins with an initial hot model, and then cools down to a specified temperature. The equation of state (EOS) tables of \citet{ld74} are used in the degenerate core and the EOS tables of \citet*{sd95} are used in the partially ionized envelope. The updated OPAL opacities of \citet{ic96} are used. The mixing length theory (MLT) of \citet{bk71} is adopted to deal with the convection. We take the MLT parameter as $\alpha =$ 0.6, which is recommended by \citet{bp95}.

Unlike our earlier work \citep{sj10}, we use the Modules for Experiments in Stellar Astrophysics \citep[MESA,][]{pb11,pb13} to generate the initial white dwarf models in this work. The MESA produces a white dwarf from a zero-age main sequence model with metallicity $Z=$ 0.02, and then evolves it through the main sequence, red giant branch and asymptotic giant branch stages, until the star becomes a hot white dwarf. All of the models are assumed to have carbon/oxygen core composition.

We incorporate the scheme of \citet*{ta94} into the WDEC to treat the element diffusion. Here, we only take four elements into account, i.e. hydrogen (H), helium (He), carbon (C) and oxygen (O). The composition profiles are then determined by the diffused results, which are different from our previous work of treating the H/He and He/C transition zones with approximations of the equilibrium profiles.

\begin{figure*}
\centering
\begin{minipage}{1.0\textwidth}
\centering
\subfloat[]{
\label{cp}
\includegraphics[scale=1.0]{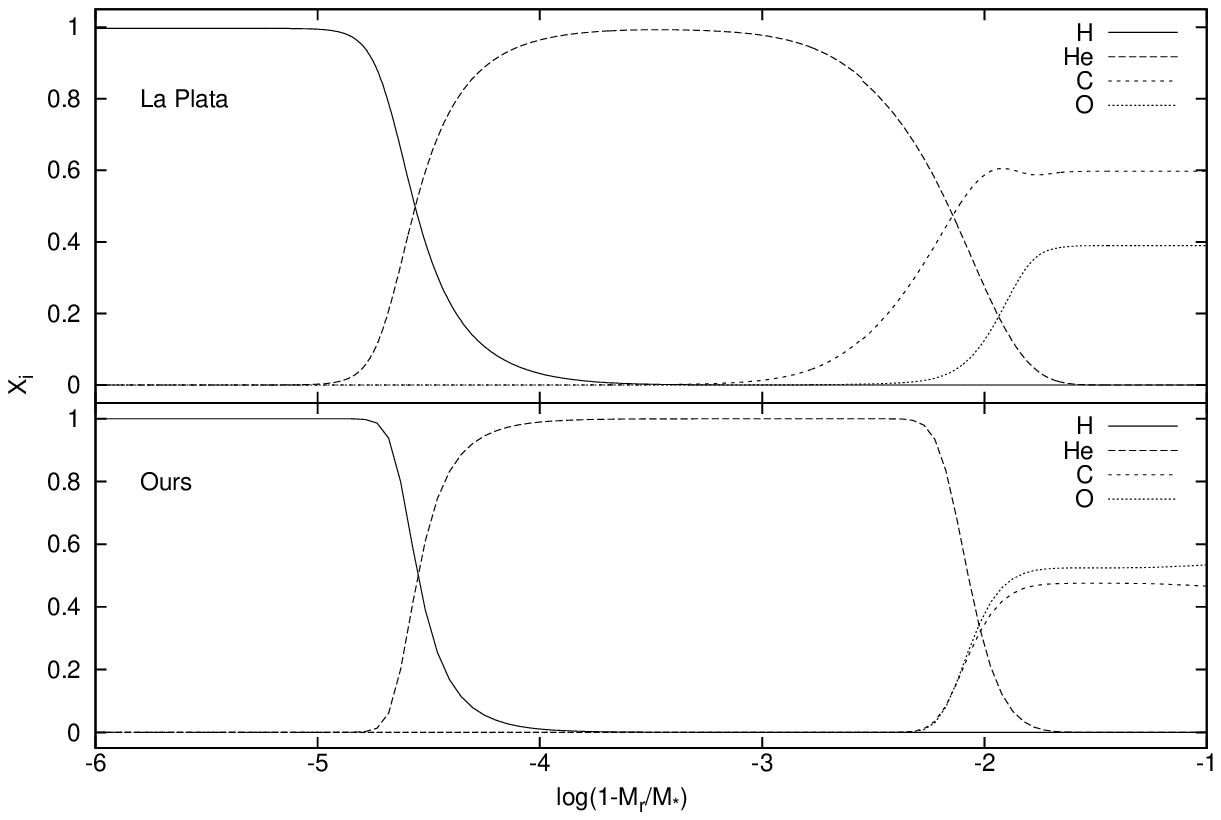}
}
\end{minipage}
\begin{minipage}{1.0\textwidth}
\centering
\subfloat[]{
\label{bn}
\includegraphics[scale=1.0]{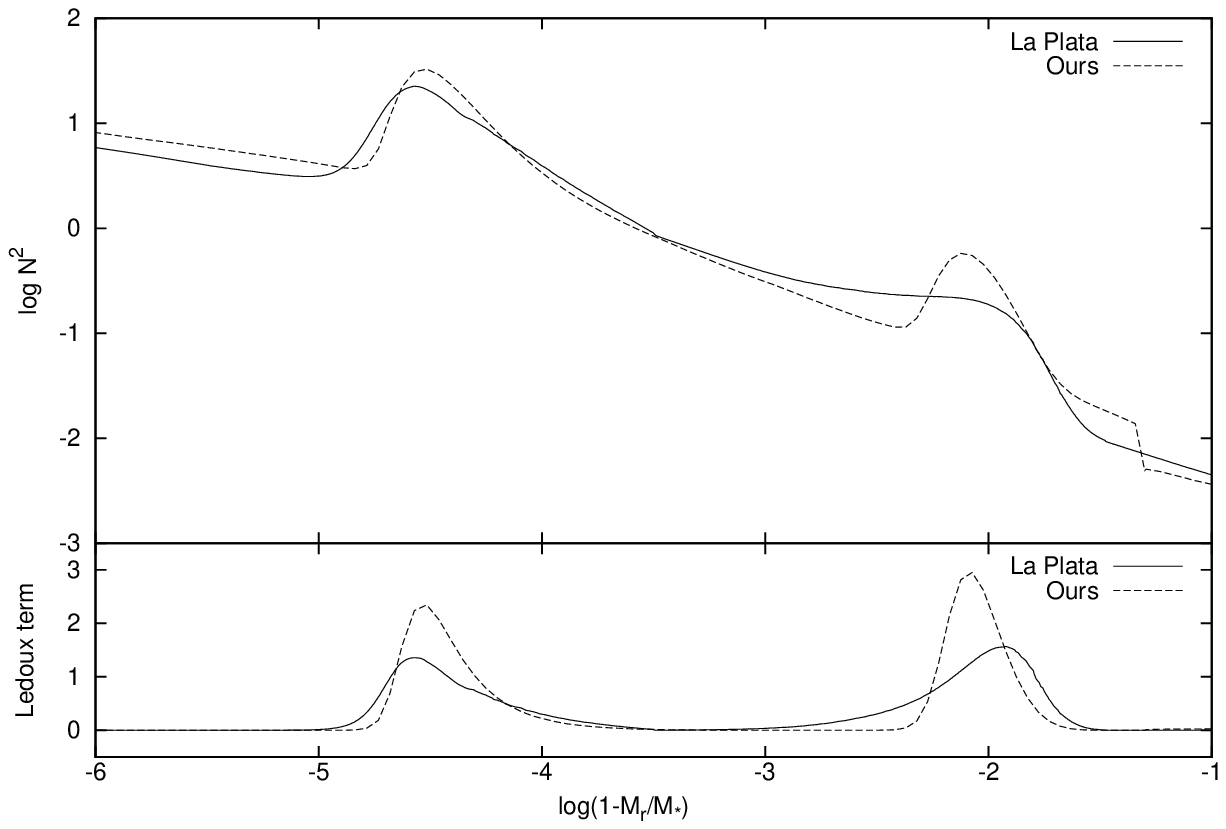}
}
\end{minipage}
\caption{(a) Comparison of the chemical profiles through the transition regions of the La Plata's model (upper panel) and our model (bottom panel). (b) Comparison of the logarithm of the squared Brunt-V\"ais\"al\"a frequency $\log N^2$ (upper panel) and the Ledoux term (bottom panel) of the La Plata's model and our model. These models are of $M_{\ast}/{\rm M_{\sun}}=0.705$, $\log(M_{\rm H}/M_{\ast})=-4.45$ and $T_{\rm eff}=11306$\,K.}
\label{we}  
\end{figure*}

We would like to compare the composition profiles through the transition regions of our models with those obtained by the La Plata group, which also compute full evolutionary sequences including the gravitational settling \citep{al10,ra12}. We choose a model of $M_{\ast}/{\rm M_{\sun}}=0.705$ and $\log(M_{\rm H}/M_{\ast})=-4.45$ from the La Plata's database and compute a homologous model of ours with the same $M_{\ast}$ and $M_{\rm H}$. The chemical profiles of those two models at $T_{\rm eff}=11306$\,K are compared in Fig.~\ref{cp}. Since the chemical transition zones have a strong impact on the Ledoux term and the Brunt-V\"ais\"al\"a frequency, they are also compared in Fig.~\ref{bn}.

We also integrate a modified pulsation code of \citet {ly92a,ly92b} into the WDEC. It calculates pulsation periods of the evolved model as soon as the evolution is done.

\subsection{Model matching} \label{matching}

The only information that we can get directly from the time-series photometric observations is the the observed periods. We calculate the $\chi^2$ value for each model using the equation
\begin{equation}
\chi^2=\frac{1}{N}\sum_{i=1}^{N}(P_{{\rm c},i}-P_{{\rm o},i})^2,
\end{equation}
where $P_{{\rm c},i}$ are pulsation periods calculated with the model, $P_{{\rm o},i}$ are observed periods and $N$ is the total number of observed periods. We only calculate pulsation periods with $l=1$ and $l=2$. Higher $l$ values are neglected, because the geometrical cancellation makes these modes hardly to be observed photometrically \citep[see][and references therein]{cb08}. Since there is a lack of previous asteroseismological study for KUV 11370+4222, it brings some indeterminacy to the confirmation of $l$ values. Fortunately, we have found a triplet, which helps us to identify the period of 398.09\,s (2511.973\,$\umu$Hz) as an $l=1$ mode. For other observed modes, we try to fit each of them to the calculated modes of either $l=1$ or $l=2$ respectively. Then we choose the calculated modes which are closest to the observed ones to calculate $\chi^2$ value. All calculated modes are of $m=0$. We thus assumed implicitly that the observed modes are all $m=0$. However, this assumption is somewhat suspicious, since some of the observed periods might not be really $m=0$ modes. But we do not have enough evidence to confirm their $m$ values, except for the triplet. It should be aware that this would bring some uncertainty to the analysis results.

The pulsation periods of the model depend on the model's structure, which is governed by four adjustable parameters: 
\begin{description}
\item[$p_{1} = M_{\ast}/{\rm M_{\sun}}$] the total mass of the white dwarf in unit of the solar mass.
\item[$p_{2} = T_{\rm eff}$] the effective temperature.
\item[$p_{3} = \log(M_{\rm He}/M_{\ast})$] the logarithmic helium mass fraction.
\item[$p_{4} = \log(M_{\rm H}/M_{\ast})$] the logarithmic hydrogen mass fraction.
\end{description} 
The ranges of parameter are set as
\begin{description}
\item[$0.5 \le p_{1} \le 0.8$] with a step of $\Delta p_{1} = 0.005$.
\item[$10850\,{\rm K} \le p_{2} \le 12250\,{\rm K}$] with a step of $\Delta p_{2} = 50\,{\rm K}$.
\item[$-4 \le p_{3} \le -2$] with a step of $\Delta p_{3} = 0.1$.
\item[$-10 \le p_{4} \le -4$] with a step of $\Delta p_{4} = 0.1$.
\end{description}
The mass range covers the most part of white dwarfs. The effective temperature range corresponds to the instability strip. The ranges of the hydrogen and helium mass fraction are considered as the typical values of the DA white dwarfs. These parameters constitute a four-dimensional (4D) parameter space.

Our goal is to find out a unique set of parameters $({\hat p}_{1},{\hat p}_{2},{\hat p}_{3},{\hat p}_{4})$, which make the model's $\chi^2$ minimum. The model is then regarded as the best-fitting model.

\subsection{Best-fitting model} \label{best-fitting}

Traversing the parameter space is a time-consuming job, especially in the multi-parameter case. In order to let the searching more efficient and robust, we adopt a genetic algorithm based on a subroutine called PIKAIA \citep{cp95} to find the best-fitting model. The genetic algorithm inspires from the biological evolution by natural selection. It evolves the population of candidate models to the better models. The evolution is an iterative process, which starts from a population of randomly generated individuals. The population in each iteration is called a generation. In each generation, we evaluate a fitness function for each individual. The fitness function is defined as $F(p_{1},p_{2},p_{3},p_{4})=1/\chi^2$. Pairs of individuals are then randomly selected from current population, which are used to breed the new generation. The probability of a certain individual being selected is in proportion to its fitness, so better models are inherited. The new generation is then used in the next iteration. The iteration terminates when the termination criterion is satisfied.

In practice, we compute 200 models, whose parameters are chosen randomly from the 4D parameter space, as the original population. After that, it evolves genetically to generate later generations. We let this process go on until 100 generations have been produced. At that time, the properties of population have already stabilized.

We finally get the best-fitting model, whose parameters are $M_{\ast}/{\rm M_{\sun}}=0.625$, $T_{\rm eff}=10950$\,K, $M_{\rm He}/M_{\ast}=10^{-2.2}$ and $M_{\rm H}/M_{\ast}=10^{-4.0}$. The helium and hydrogen mass fractions approximate to the typical values of the DA white dwarfs, which are $M_{\rm He}/M_{\ast} \sim 10^{-2}$, and $M_{\rm H}/M_{\ast} \sim 10^{-4}$.

\begin{figure*}
\centering
\begin{minipage}{1.0\textwidth}
\centering
\subfloat[]{
\label{2d_a}
\includegraphics[scale=0.6]{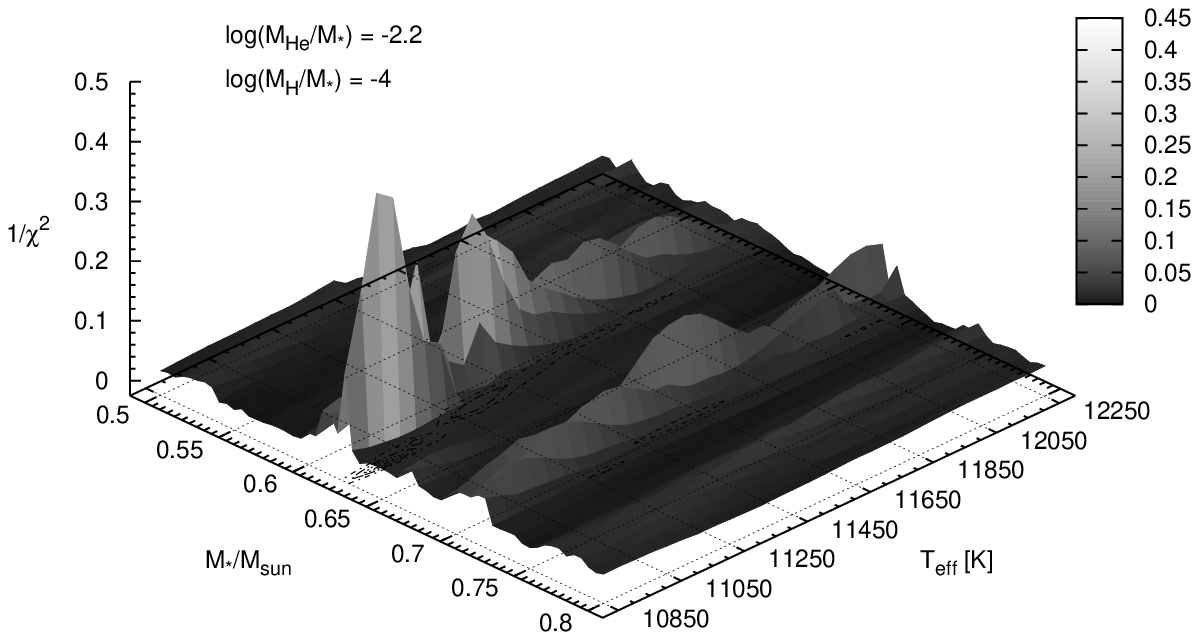}
}
\subfloat[]{
\label{2d_b}
\includegraphics[scale=0.6]{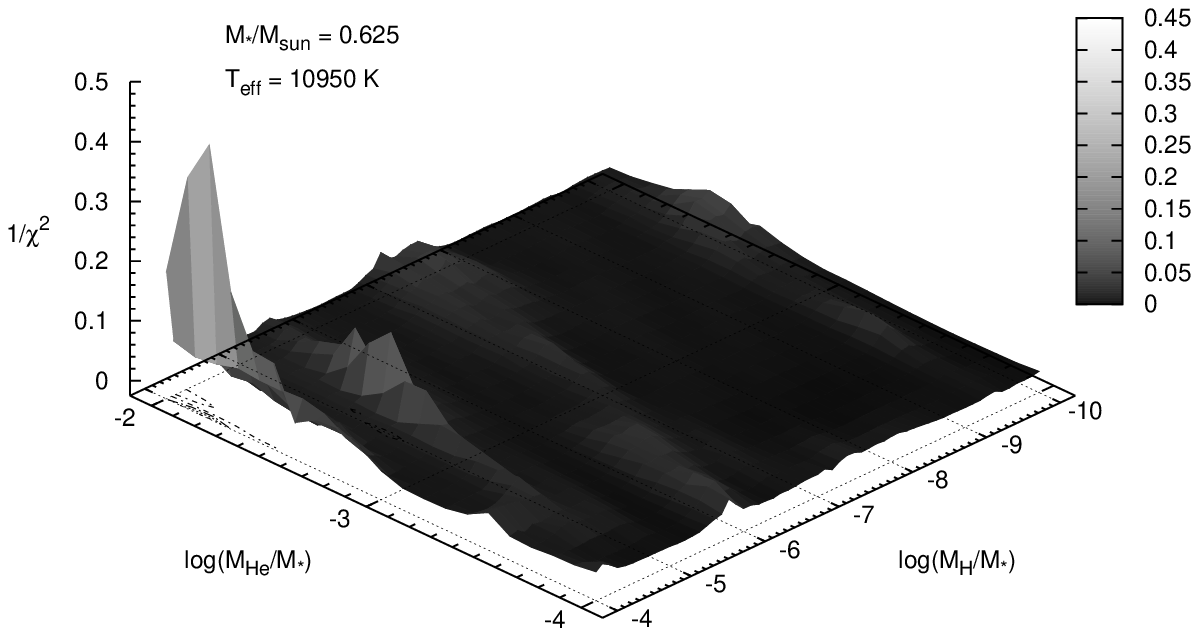}
}
\end{minipage}
\begin{minipage}{1.0\textwidth}
\centering
\subfloat[]{
\label{2d_c}
\includegraphics[scale=0.6]{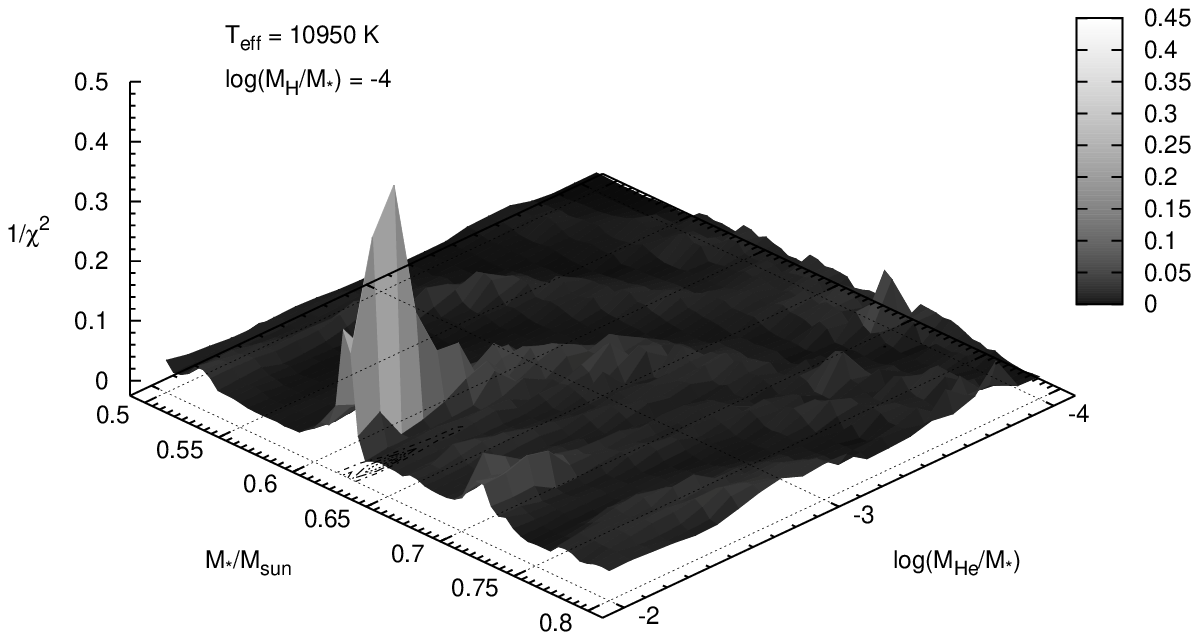}
}
\subfloat[]{
\label{2d_d}
\includegraphics[scale=0.6]{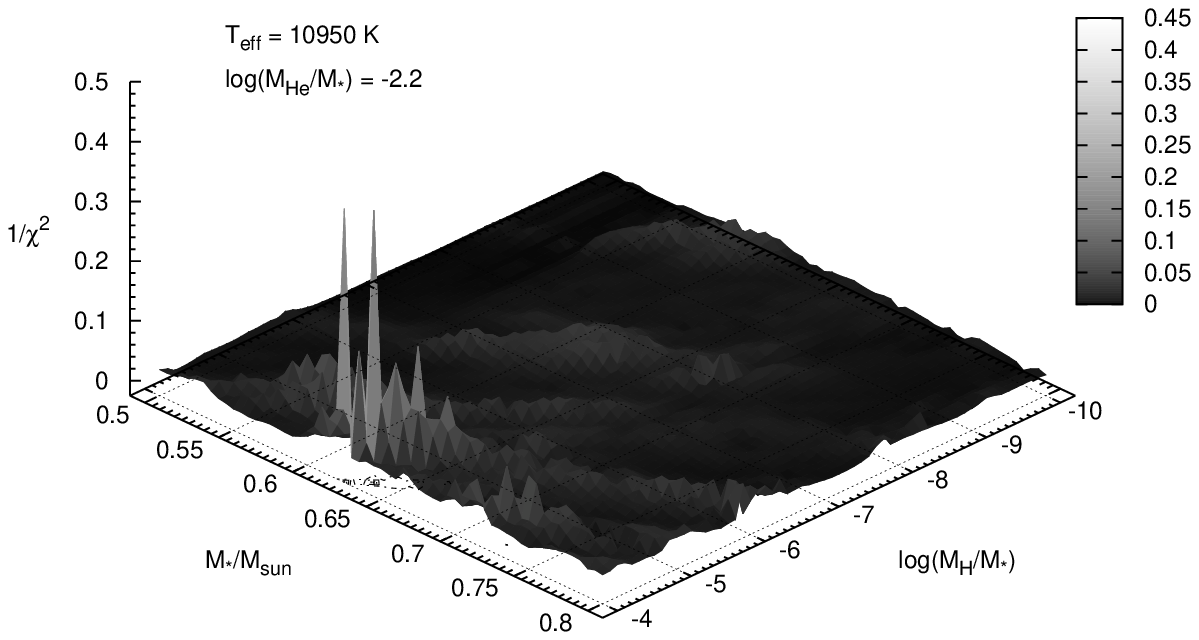}
}
\end{minipage}
\begin{minipage}{1.0\textwidth}
\centering
\subfloat[]{
\label{2d_e}
\includegraphics[scale=0.6]{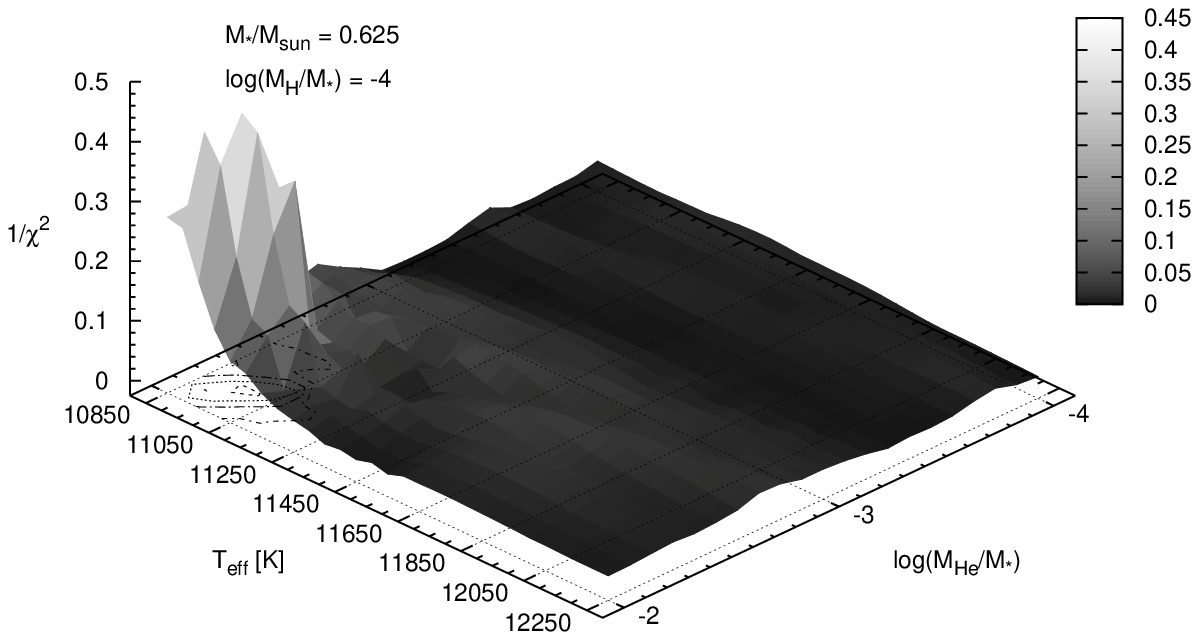}
}
\subfloat[]{
\label{2d_f}
\includegraphics[scale=0.6]{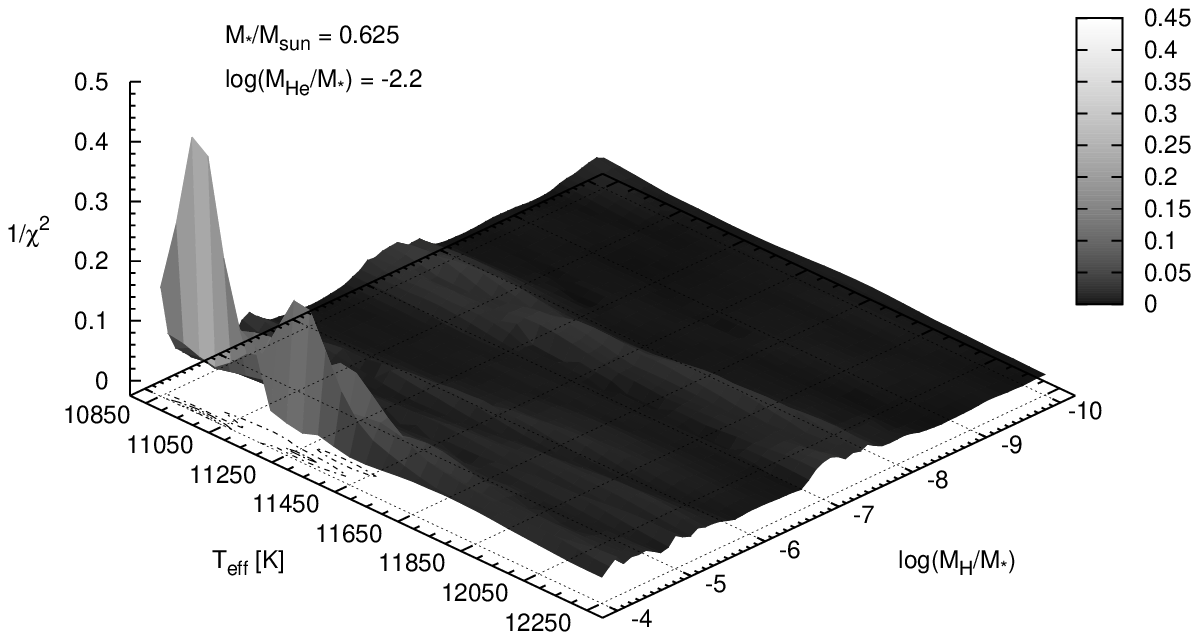}
}
\end{minipage}
\caption{3D figures that show $F(p_{1},p_{2},p_{3},p_{4})$ vary with two arbitrary parameters. (a) The plot that chooses $p_{1}$, $p_{2}$ as independent variables. (b) The plot that chooses $p_{3}$, $p_{4}$ as independent variables. (c) The plot of $F(p_{1},p_{2},p_{3},p_{4})$ versus $p_{1}$, $p_{3}$. (d) The plot of $F(p_{1},p_{2},p_{3},p_{4})$ versus $p_{1}$, $p_{4}$. (e) The plot that $p_{2}$, $p_{3}$ are used as the independent variables. (f) The plot that $p_{2}$, $p_{4}$ are used as the independent variables. Contour lines are also plotted at the base of each figure.}
\label{2d} 
\end{figure*}

In Fig.~\ref{2d} we show six 3D figures of $F(p_{1},p_{2},p_{3},p_{4})$ using two arbitrary parameters as independent variables. Each figure is a cut of the 4D parameter space, which is obtained by fixing two of the parameters at the values of the best-fitting model and varying the others. We vary $p_{1}$, $p_{2}$ while fixing $p_{3}$, $p_{4}$ to get Fig.~\ref{2d_a}. Contrarily, we vary $p_{3}$, $p_{4}$ while fixing $p_{1}$, $p_{1}$ thus get Fig.~\ref{2d_b}. We plot $F(p_{1},p_{2},p_{3},p_{4})$ versus $p_{1}$, $p_{3}$ in Fig.~\ref{2d_c} and plot $F(p_{1},p_{2},p_{3},p_{4})$ versus $p_{1}$, $p_{4}$ in Fig.~\ref{2d_d}. In the last two figures, $p_{2}$, $p_{3}$ and $p_{2}$, $p_{4}$ are used as the independent variables, respectively. We also plot contour lines at the base of each figure. The best-fitting model is distinguished by the peaks in these figures, which correspond to $F(p_{1},p_{2},p_{3},p_{4}) = 0.433$ and hence $\chi^2=2.31$.

It should be noticed that the mass and the effective temperature of our best-fitting model are significantly different from the spectroscopic result ($M_{\ast}/{\rm M_{\sun}}=0.74$, $T_{\rm eff}=12230$\,K) of \citet{ga11}, and also from the best-fitting model parameters ($M_{\ast}/{\rm M_{\sun}}=0.632$, $T_{\rm eff}=11237$\,K) derived from asteroseismology by \citet{ra12}. The analysis of \citeauthor{ra12} was based on the only three frequencies found in the discovery observations in 1996. The present analysis which uses the newly discovered frequencies gives a different result. This is worth emphasizing. Moreover, the differences may reflect some differences between the models of \citeauthor{ra12} and ours.

\subsection {Mode identification} \label{identification}

A part of the calculated periods ($P_{\rm c}$) of the best-fitting model are listed in Table~\ref{periods}. The observed periods ($P_{\rm o}$) are listed after the corresponding calculated periods. We also list the $(P_{\rm c}-P_{\rm o})^2$ values in the table.

We hence identify the $l$ and $n$ values of the observed modes by comparing them to the calculated modes. The triplet (394.14, 398.09 and 402.06\,s) is identified as $l=1$ in Section \ref{triplet}. Here they are identified as $n=6$ modes. Another $l=1$ mode is the 280.80\,s, which is of $n=4$. The other six observed modes: 139.03, 257.88, 293.77, 462.83, 762.42 and 811.75\,s are identified as $l=2$. Their $n$ values are 3, 7, 9, 16, 28 and 30, respectively.

\begin{table*}
\centering
\caption{Part of calculated periods ($P_{\rm c}$) of the best-fitting model. The calculated periods are listed with their $l$ and $n$ values. The observed periods ($P_{\rm o}$) are listed beside the corresponding calculated periods and the values of $(P_{\rm c}-P_{\rm o})^2$ are also given.}
\begin{tabular}{cccccccc}
\hline               
\multicolumn{4}{l}{$l=1$} & \multicolumn{4}{l}{$l=2$} \\
$n$ & $P_{\rm c}$ (s) & $P_{\rm o}$ (s) & $(P_{\rm c}-P_{\rm o})^2$ & $n$ & $P_{\rm c}$ (s) & $P_{\rm o}$ (s) & $(P_{\rm c}-P_{\rm o})^2$ \\
\hline
 1 &  107.44 &        &      &  1 &  62.07 &        &      \\
 2 &  208.12 &        &      &  2 & 120.41 &        &      \\
 3 &  237.73 &        &      &  3 & 138.06 & 139.03 & 0.95 \\
 4 &  279.67 & 280.80 & 1.27 &  4 & 161.73 &        &      \\
 5 &  338.13 &        &      &  5 & 195.37 &        &      \\
 6 &  396.57 & 398.09 & 2.31 &  6 & 230.81 &        &      \\
 7 &  408.85 &        &      &  7 & 260.59 & 257.88 & 7.32 \\
 8 &  454.52 &        &      &  8 & 276.14 &        &      \\
 9 &  505.69 &        &      &  9 & 294.53 & 293.77 & 0.58 \\
10 &  549.09 &        &      & 10 & 319.09 &        &      \\
11 &  596.13 &        &      & 11 & 347.04 &        &      \\
12 &  624.54 &        &      & 12 & 375.09 &        &      \\
13 &  653.95 &        &      & 13 & 403.16 &        &      \\
14 &  701.97 &        &      & 14 & 425.88 &        &      \\
15 &  750.32 &        &      & 15 & 441.74 &        &      \\
16 &  796.68 &        &      & 16 & 464.84 & 462.83 & 4.05 \\
17 &  839.58 &        &      & 17 & 491.55 &        &      \\
18 &  874.85 &        &      & 18 & 518.12 &        &      \\
19 &  913.45 &        &      & 19 & 539.48 &        &      \\
20 &  960.57 &        &      & 20 & 560.19 &        &      \\
21 & 1008.56 &        &      & 21 & 586.59 &        &      \\
22 & 1052.52 &        &      & 22 & 614.40 &        &      \\
23 & 1086.51 &        &      & 23 & 641.48 &        &      \\
24 & 1120.88 &        &      & 24 & 668.92 &        &      \\
25 & 1165.44 &        &      & 25 & 695.29 &        &      \\
26 & 1212.89 &        &      & 26 & 716.37 &        &      \\
27 & 1260.40 &        &      & 27 & 736.80 &        &      \\
28 & 1305.54 &        &      & 28 & 761.89 & 762.42 & 0.28 \\
29 & 1349.36 &        &      & 29 & 787.58 &        &      \\
30 & 1394.08 &        &      & 30 & 813.06 & 811.75 & 1.72 \\
\hline
\end{tabular}
\label{periods}
\end{table*}

In the previous work of \citet{ra12} the three modes found in 1996 had been identified. Here we compare them in terms of $l$ and $n$ with our results in Table~\ref{results}. It also shows remarkable differences between the results.

\begin{table*}
\centering
\caption{Comparison of the results of mode identification between \citeauthor{ra12} and present work. We list the periods ($P$) of these modes as well as their $l$ and $n$ values.}
\begin{tabular}{cccccc}
\hline               
\multicolumn{3}{c}{\citeauthor{ra12}} & \multicolumn{3}{c}{Present work} \\
$P$ (s) & $l$ & $n$ & $P$ (s) & $l$ & $n$ \\
\hline
257.20 & 1 &  3 & 257.88 & 2 &  7 \\
292.20 & 1 &  4 & 293.77 & 2 &  9 \\
462.90 & 2 & 15 & 462.83 & 2 & 16 \\
\hline
\end{tabular}
\label{results}
\end{table*}

\subsection {Rotation period} \label{rotation}

We have considered the three frequencies 2487.19, 2511.97 and 2537.14\,$\umu$Hz as a triplet, which is due to the splitting of an $l=1$ mode by rotation. In the existence of rotation, the frequency of an $l,n$ mode $\omega_{l,n}$ is changed to
\begin{equation}
\omega_{l,n,m} = \omega_{l,n} + m (1-C_{l,n})\Omega,
\end{equation}
where $\Omega$ is the rotation frequency and $C_{l,n}$ comes from the Coriolis force term in the momentum equation. In the asymptotic approximation, we have 
\[
C_{l,n} = \frac{1}{l(l+1)}.
\]
The average value of the frequency separation is $24.98\pm0.38$\,$\umu$Hz and $|m|=1$. So we derive the rotation period $P_{\rm rot} = 5.56\pm0.08$\,h. It implies a fast rotation rate of star.

\subsection {Mode trapping} \label{trapping}

When a mode has a node of its eigenfunction in the chemical transition zones, i.e. the thickness of the hydrogen (or helium) layer is multiple of the wavelength of the mode, there will be a resonance formed in that layer. It causes the oscillation to concentrate in the outer layer and to become weaker in the core. The mode is hence trapped in the outer layer, so this mechanism is called mode trapping.

\begin{figure*}
\centering
\begin{minipage}{1.0\textwidth}
\centering
\subfloat[]{
\label{dp_a}
\includegraphics[scale=1.0]{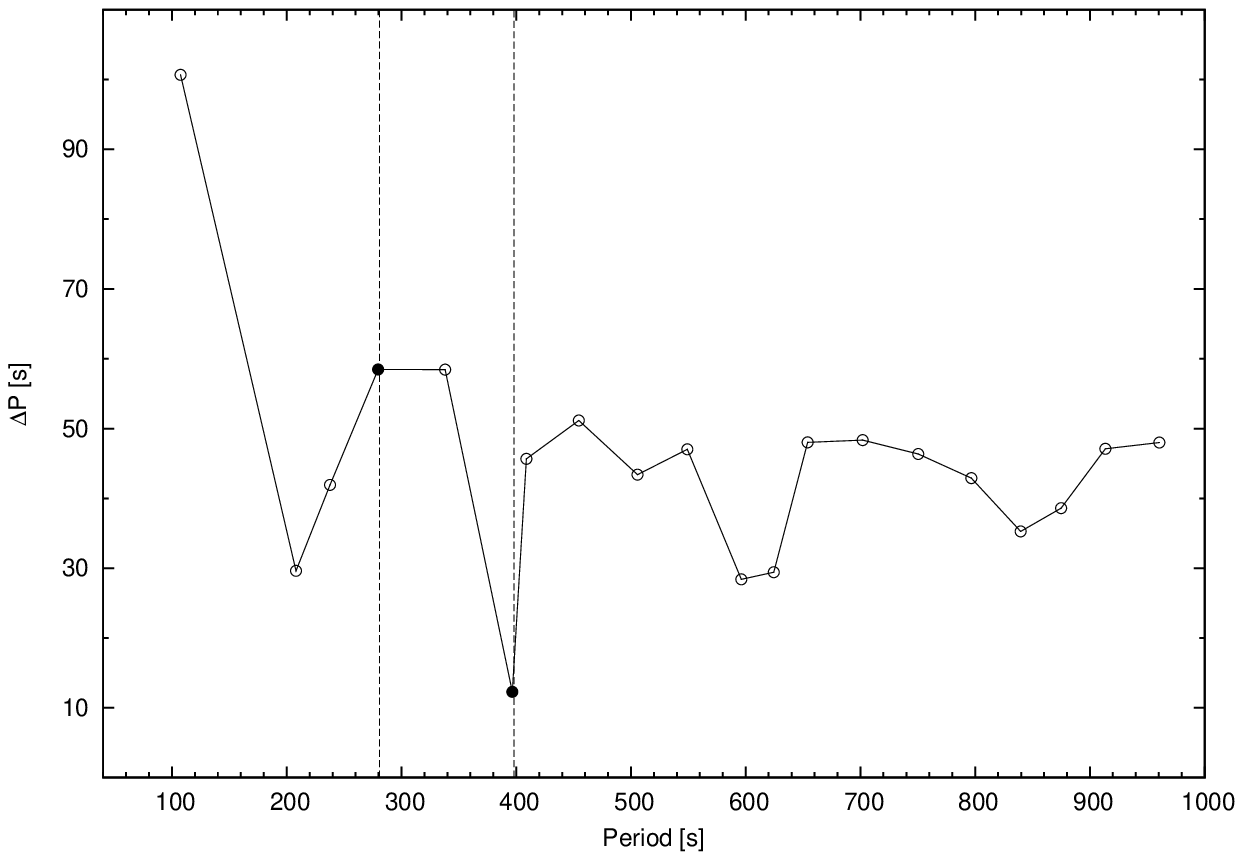}
}
\end{minipage}

\begin{minipage}{1.0\textwidth}
\centering
\subfloat[]{
\label{dp_b}
\includegraphics[scale=1.0]{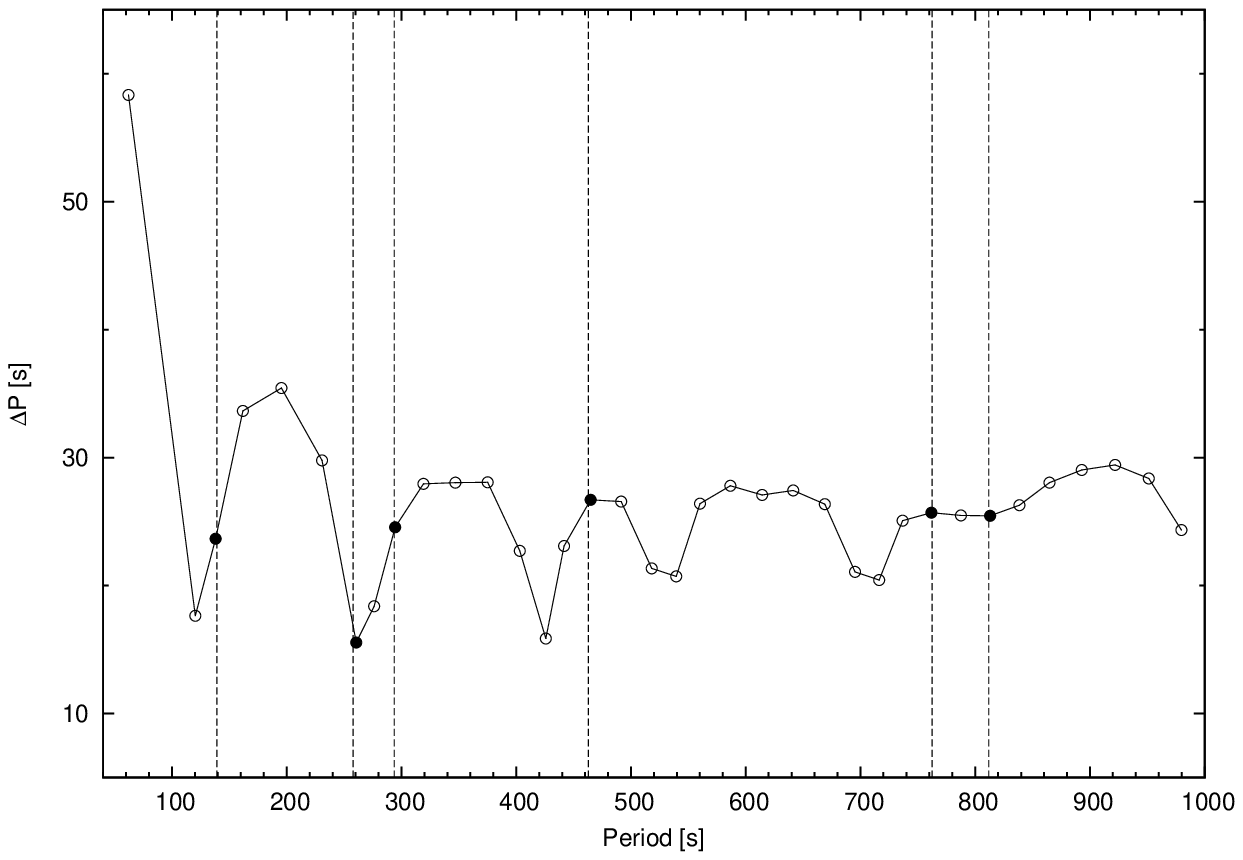}
}
\end{minipage}
\caption{Diagrams show deviations of the model periods from the the uniform period spacing. (a) Plot of $l=1$ modes. (b) Same as (a) but for $l=2$ modes. $\Delta P$ are period spacing of consecutive modes. Each data point represents a calculated period of the model. The vertical dashed lines denote the observed periods. Calculated periods who are used to fit the observed ones are denoted with filled circles. The minima in the diagrams are considered as trapped modes.}
\label{dp} 
\end{figure*}

The kinetic energy of a pulsation mode is defined as
\begin{equation} \label{ek}
E_{\rm kin}=2\pi\omega^2\int_{0}^{R}[\xi_{\rm r}^2+l(l+1)\xi_{\rm h}^2]\rho r^2 dr,
\end{equation}
where $\omega$ is the frequency, $\xi_{\rm r}$ is the radial displacement, $\xi_{\rm h}$ is the horizontal displacement, $\rho$ is the density and $r$ is the radius. The integrand is defined as kinetic energy weight function. The major contribution to the kinetic energy comes from the core, since the density of the core is greatly higher than the outer layer for white dwarf stars. The trapped modes will have lower kinetic energy because of their weak amplitudes in the core. Moreover, the mode trapping will cause the pulsation periods to deviate from the uniform period spacing. It is shown in Fig.~\ref{dp}. $\Delta P=P_{l,n+1}-P_{l,n}$ are the period spacing of consecutive modes. Each data point represents a calculated period of the model. The vertical dashed lines denote the observed periods. Those calculated periods who are used to fit the observed ones are denoted with filled circles. The minima in the period spacing diagrams correspond to the modes whose kinetic energy are also minima \citep{bp91}, and they are considered as trapped modes.

\begin{figure*}
\centering
\begin{minipage}{1.0\textwidth}
\centering
\subfloat[]{
\label{we_a}
\includegraphics[scale=1.0]{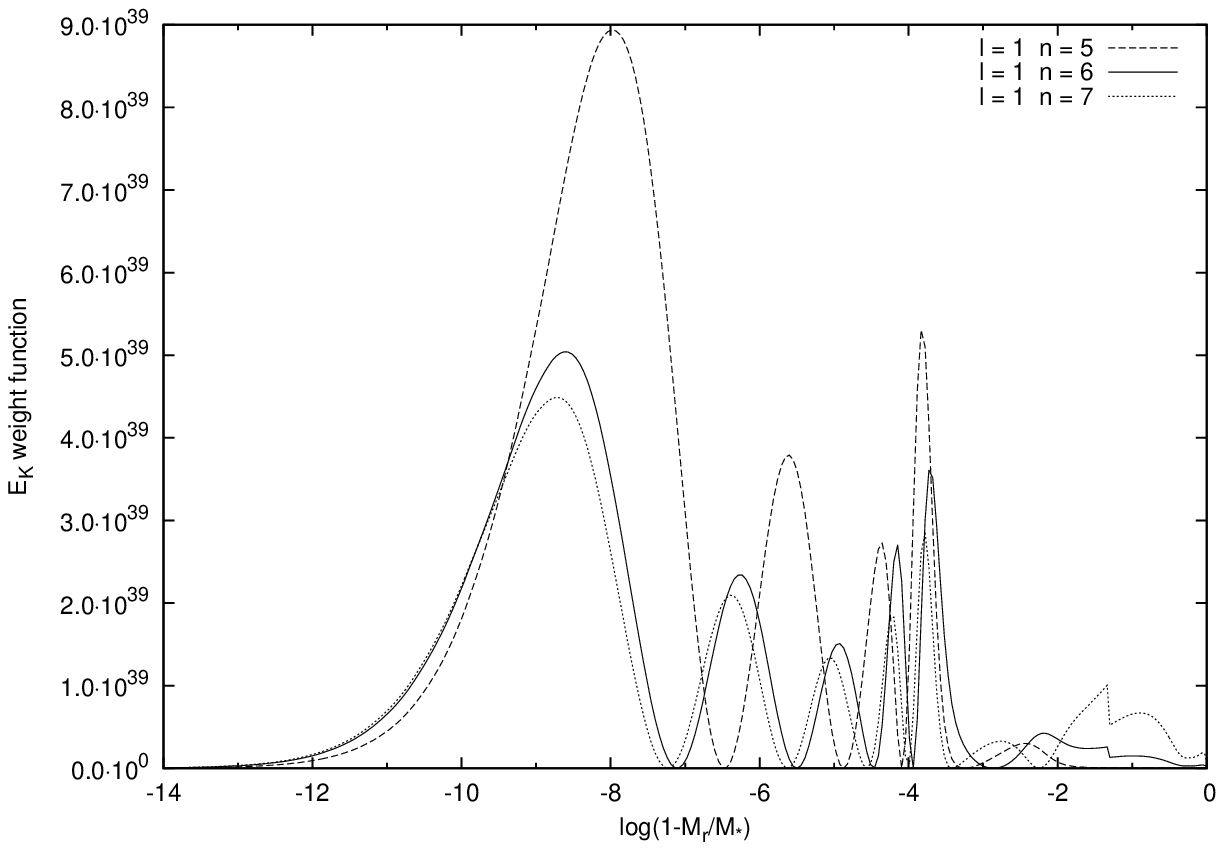}
}
\end{minipage}
\begin{minipage}{1.0\textwidth}
\centering
\subfloat[]{
\label{we_b}
\includegraphics[scale=1.0]{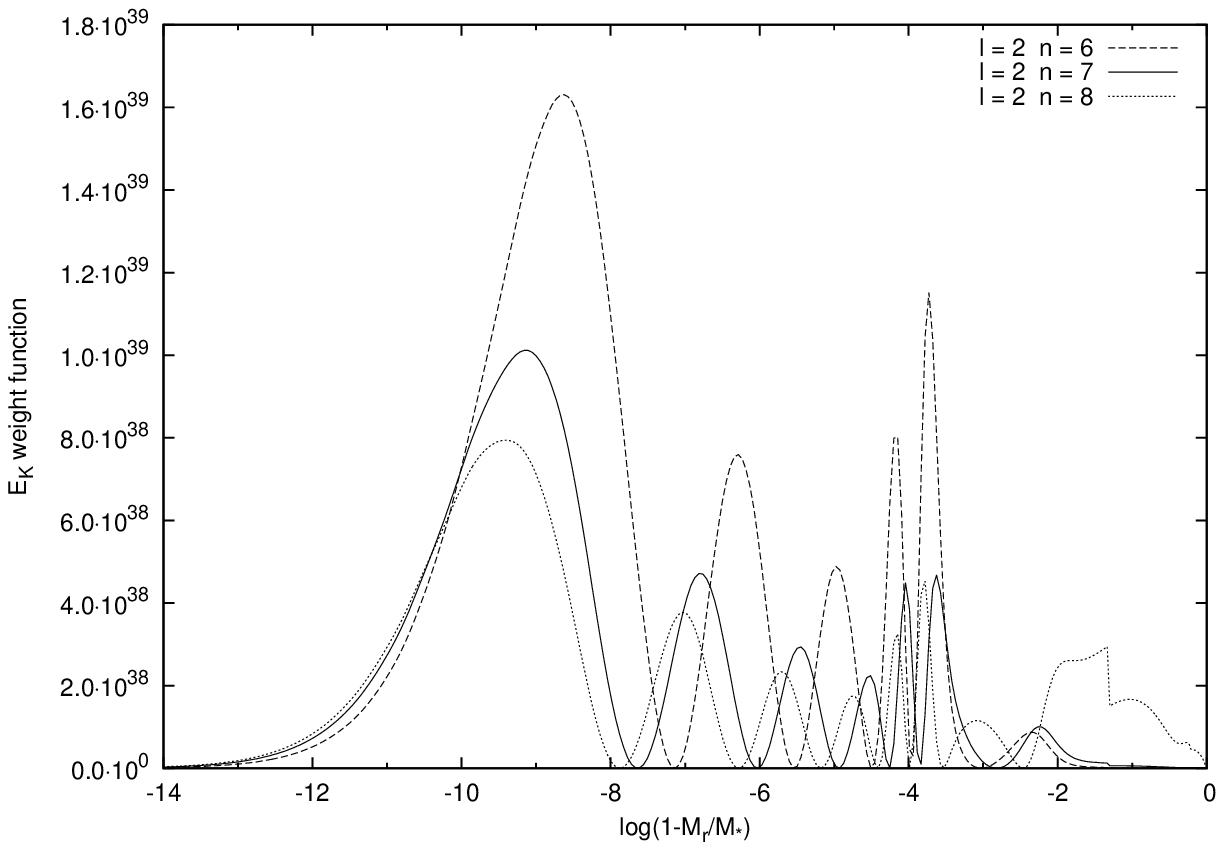}
}
\end{minipage}
\caption{(a) Plot of the kinetic energy weight function for three consecutive $l=1$ modes: $n=5$ (dashed line), $n=6$ (solid line) and $n=7$ (dotted line). (b) Same as (a) but for $l=2$ modes: $n=6$ (dashed line), $n=7$ (solid line) and $n=8$ (dotted line).}
\label{we} 
\end{figure*}

Two of the observed modes: 398.09 and 257.88\,s are considered as trapped modes since they correspond to the minima in Fig.~\ref{dp}. The corresponding calculated modes are 396.57 ($l=1,n=6$) and 260.59\,s ($l=2,n=7$). To see the distribution of kinetic energy in the interior of the model, we plot the kinetic energy weight function (the integrand of Eq.~\ref{ek}) of the two calculated modes as well as those of their neighbours in Fig.~\ref{we_a} and Fig.~\ref{we_b}, respectively. The kinetic energy weight function of the $l=1,n=6$ mode shows a lower amplitude throughout the model. The kinetic energy of this mode would be less than its neighbours, whose kinetic energy weight function have considerably larger amplitudes either in the outer layer (the $l=1,n=5$ mode) or in the core (the $l=1,n=7$ mode). The case of the $l=2,n=7$ mode is similar. The logarithmic kinetic energy ($\log E_{\rm kin}$) of these modes are listed in Table~\ref{logek}. It shows that the kinetic energy of the $l=1,n=6$ and $l=2,n=7$ modes are indeed lower than their neighbours, which emphasize the conclusion that the two modes are trapped modes.

\begin{table*}
\centering
\caption{The logarithms of kinetic energy ($\log E_{\rm kin}$) of the two trapped modes and their neighbours.}
\begin{tabular}{cccccccc}
\hline               
$l$ & $n$ & Period (s) & $\log E_{\rm kin}$ & $l$ & $n$ & Period (s) & $\log E_{\rm kin}$ \\
\hline
1 & 5 & 338.13 & 44.53 & 2 & 6 & 230.81 & 44.15 \\
1 & 6 & 396.57 & 44.36 & 2 & 7 & 260.59 & 43.84 \\
1 & 7 & 408.85 & 44.61 & 2 & 8 & 276.14 & 44.31 \\
\hline
\end{tabular}
\label{logek}
\end{table*}

The trapped modes are generally expected to have higher amplitudes since they have higher growth rates \citep{bp91}. It is true for the 257.88\,s mode, which appears as the largest amplitude mode both in the data of 1996 and 2010 (see Table~\ref{compare}), just corresponds with a trapped mode. Other observed periods are often found to be near to the minima in Fig.~\ref{dp}. This is also similar to the conclusion of \citet{bz09}.

\section{Summary and conclusions} \label{summary}

The observations and model analyses of the ZZ Ceti star KUV 11370+4222 are presented above. The results are summarised as below:
\begin{enumerate}
\item We have obtained eight nights' photometric data for KUV 11370+4222 in 2010. This is the first observation run to this object after it was found as a ZZ Ceti star in 1996. We obtain twelve frequencies of S/N\,$>$\,4 from the FT spectrum of the light curves. Ten of them are recognized as independent modes. The remainders are one harmonic and one linear combination.
\item One triplet is found, which reflects the frequency splitting of $l=1$ mode due to rotation of the star. We get the average value of the frequency separation in the triplet as $24.98\pm0.38$\,$\umu$Hz and thus derive an average rotation period of $5.56\pm0.08$\,h.
\item Matching the calculated periods to the observed ones, we find the best-fitting model whose essential parameters are $M_{\ast}/{\rm M_{\sun}} = 0.625$, $T_{\rm eff}=10950$\,K, $M_{\rm He}/M_{\ast}=10^{-2.2}$ and $M_{\rm H}/M_{\ast}=10^{-4.0}$. 
\item By comparing with the calculated modes, we identify $l$ and $n$ values of the observed modes. The triplet and $f=3561.21$\,$\umu$Hz mode are identified as $l=1,n=6$ and $l=1,n=4$ modes, respectively. The other six observed modes are identified as $l=2$ modes. Their $n$ values are 3, 7, 9, 16, 28 and 30, respectively. 
\item We investigate the property of the mode trapping and find that the largest amplitude mode is just a trapped mode.
\end{enumerate}

Further observations of KUV 11370+4222 are needed in the future, which would help one to find more independent modes and multiplets. It would give better constraints on the modelling of KUV 11370+4222.
 
\section*{Acknowledgments}

JS and YL acknowledge the support from the Knowledge Innovation Program of the Chinese Academy of Sciences through Grant KJCX2-YW-T24, and the partial support from the National Natural Science Foundation of China (11273054). JNF and CL acknowledge the support from the Joint Fund of Astronomy of National Natural Science Foundation of China (NSFC) and Chinese Academy of Sciences through the Grant U1231202, and the support from the National Basic Research Program of China (973 Program 2014CB845700 and 2013CB834900).

\label{lastpage}

\end{document}